\documentclass[aps,prc,twocolumn,epsfig,showpacs,floatfix]{revtex4-1}
\usepackage{lineno}
\usepackage{epsfig}
\usepackage{graphicx}
\usepackage{dcolumn}
\usepackage[squaren,Gray]{SIunits}
\usepackage[colorlinks=true, linkcolor=blue, citecolor=blue, urlcolor=black]{hyperref}
\usepackage{comment}

\begin{document}

\title{Microscopic study of  $^{40}$Ca+$^{58,64}$Ni fusion reactions}

\author{D. Bourgin$^{1,2}$}
\email{dominique.bourgin@iphc.cnrs.fr}
\author{C. Simenel$^3$}
\author{S. Courtin$^{1,2,4}$}
\author{F. Haas$^{1,2}$}

\affiliation{$^1$IPHC, Universit\'e de Strasbourg, F-67037 Strasbourg, France}
\affiliation{$^2$CNRS, UMR7178, F-67037 Strasbourg, France}
\affiliation{$^3$Department of Nuclear Physics, RSPE, Australian National University, Canberra, Australian Capital Territory 0200, Australia}
\affiliation{$^4$USIAS, F-67083 Strasbourg, France}

\date{\today}

\begin{abstract}
\noindent \textbf{Background:} Heavy-ion fusion reactions at energies near the Coulomb barrier are influenced by couplings between the relative motion and nuclear intrinsic degrees of freedom of the colliding nuclei. The time-dependent Hartree-Fock (TDHF) theory,  incorporating the couplings at the mean-field level, as well as the coupled-channels (CC) method are standard approaches to describe low energy nuclear reactions. 

\noindent \textbf{Purpose:} To investigate the effect of couplings to inelastic and transfer channels on the fusion cross sections for the reactions $^{40}$Ca+$^{58}$Ni and $^{40}$Ca+$^{64}$Ni.

\noindent \textbf{Methods:} Fusion cross sections around and below the Coulomb barrier have been obtained from coupled-channels (CC) calculations, using the bare nucleus-nucleus potential calculated with the frozen Hartree-Fock method and coupling parameters taken from known nuclear structure data. The fusion thresholds and neutron transfer probabilities have been calculated with the TDHF method.

\noindent \textbf{Results:} For $^{40}$Ca+$^{58}$Ni, the TDHF fusion threshold is in agreement with the most probable barrier obtained in the CC calculations including the couplings to the low-lying octupole $3_1^{-}$ state for $^{40}$Ca and to the low-lying quadrupole $2_1^{+}$ state for $^{58}$Ni. This indicates that the octupole and quadrupole states are the dominant excitations while neutron transfer  is shown to be weak. For $^{40}$Ca+$^{64}$Ni, the TDHF barrier is lower than predicted by the CC calculations including the same inelastic couplings as those for $^{40}$Ca+$^{58}$Ni. TDHF calculations show large neutron transfer probabilities in $^{40}$Ca+$^{64}$Ni which could result in a lowering of the fusion threshold. 

\noindent \textbf{Conclusions:} Inelastic channels play an important role in $^{40}$Ca+$^{58}$Ni and $^{40}$Ca+$^{64}$Ni reactions. The role of neutron transfer channels has been highlighted in $^{40}$Ca+$^{64}$Ni.

\end{abstract}

\maketitle

\section{Introduction}

Heavy-ion fusion cross sections near the Coulomb barrier are influenced by couplings of the relative motion to nuclear shape deformations and vibrations  \cite{lei95,das98,bac14}. The effects of these couplings can be investigated by coupled-channels  (CC) calculations \cite{das83a,das83b,lin84,tho85,esb87,ste90,bal98,hag12} and time-dependent Hartree-Fock (TDHF) calculations \cite{was08,sim13b,sim13c}.

Nucleon transfer channels have also been shown to affect the fusion process \cite{ste95b,jia10,eve11,mon13,jia14}. 
However, a deep understanding of the interplay between fusion and transfer is still needed. 
This is partly due to the difficulty to construct predictive theoretical models of fusion incorporating the dissipation and Q-value effects of transfer. Transfer channels are indeed incorporated within CC models in a simplified fashion \cite{kar15}. 
In addition, recent progresses of microscopic description of transfer reactions are mostly limited to the mean-field approximation \cite{sim08,uma08a,sim10b,sek13,sek14}, to small  \cite{mar85,bon85,sim11} and semi-classical \cite{was09b,yil11,yil14} fluctuations. Nevertheless, comparison between experimental data and theoretical calculations of fusion properties such as barriers and cross sections is expected to shed light on the importance of coupling to transfer channels. 

On the one hand, CC calculations treat the couplings to inelastic channels in a fully quantum mechanical approach. 
However, they require input parameters for the nuclear potential as well as coupling parameters of collective states in the colliding nuclei. The TDHF approach, on the other hand, often uses  Skyrme energy density functionals \cite{sky56} which are fitted on basic nuclear structure properties \cite{vau72} as a unique input. It also naturally incorporates both coupling to inelastic and transfer channels.  However, these couplings are only treated at the mean-field level. A new approach has been proposed to combine these two complementary theories in Ref.~\cite{sim13c}. Other works have also incorporated microscopic ingredients in CC calculations \cite{hag12}.

Recently, the fusion excitation functions of $^{40}$Ca+$^{58}$Ni and $^{40}$Ca+$^{64}$Ni \cite{bou14} have been measured at energies around and below the Coulomb barrier to study the influence of the projectile and target nuclear structures on the fusion process. CC calculations were performed with the \textsc{ccfull} code \cite{hag99}, using the Aky\"uz-Winther nuclear potential \cite{aky81,win95} and including the most relevant inelastic channels, i.e. the octupole-phonon excitation ($0_{\rm gs}^+ \rightarrow 3_{1}^{-}$) for $^{40}$Ca and the quadrupole-phonon excitation ($0_{\rm gs}^+ \rightarrow 2_{1}^{+}$) for both $^{58}$Ni and $^{64}$Ni. The effect of the positive $Q$ value two-neutron transfer channel in $^{40}$Ca+$^{64}$Ni was also schematically taken into account. The $^{40}$Ca+$^{64}$Ni system has positive nucleon transfer $Q$ values, whereas $^{40}$Ca+$^{58}$Ni has only negative nucleon transfer $Q$ values. The fusion cross sections for $^{40}$Ca+$^{58}$Ni are well reproduced at sub-barrier energies by including couplings to inelastic channels in the CC calculations. For $^{40}$Ca+$^{64}$Ni, the fusion cross sections are underestimated by including only couplings to inelastic channels. The additional coupling to the two-neutron transfer channel turned out to be essential to describe the large fusion cross sections at sub-barrier energies for this system \cite{bou14}.
\newline

In this work, we have analysed the two fusion reactions $^{40}$Ca+$^{58}$Ni and $^{40}$Ca+$^{64}$Ni with microscopic calculations, employing the Skyrme-functionnal SLy4$d$ \cite{kim97}. The method is  described in Sec.~ \ref{section_TDHF_method}. Results of the calculations are presented in Sec.~\ref{section_Results} and compared with experimental data. In Sec.~\ref{section_Summary}, we summarize our conclusions.

\section{Theoretical approach}
\label{section_TDHF_method}

The time-dependent Hartree-Fock (TDHF) theory is a mean-field approximation of the many-body dynamics. 
It was initially proposed by Dirac \cite{dir30} to describe electrons in atoms, in which case the particles interact via the Coulomb interaction. Direct computation of the Hartree-Fock mean-field in nuclei remains a problem as the interaction between nucleons in the nuclear medium is still unknown. However, the development of phenomenological Skyrme effective interactions \cite{sky56} allowed to bypass the problem by using the resulting energy density functional (instead of the nucleon-nucleon interaction) to express the mean-field potential \cite{vau72}. 

Many parameterisations of the Skyrme functional have been developed over the past. However, most of the Skyrme parameterisations have been fitted assuming a one-body center-of-mass correction to describe the nuclei in their intrinsic frame.
This is implemented by replacing the nucleon mass $m$ with $Am/(A-1)$, where $A$ is the number of nucleons. Although this usually improves the structure of light nuclei, such correction should not be applied when studying collisions as it induces a spurious dependence on the initial number of nucleons of the collision partners. Therefore, we use the SLy4$d$ parameterisation \cite{kim97} which has been derived without center-of-mass correction. Note that other recent parameterisations have been derived without this correction, such as the Quark-Meson Coupling 700 \cite{gui06} and the Universal Nuclear Energy Density Functional \cite{kor12} parameterisations. Comparison with these interactions will be the subject of a future work.

The TDHF approach has become a standard tool to describe nuclear reactions (see Refs. \cite{neg82,sim12b} for a review). 
This has been made possible thanks to calculations in three dimensions including spin-orbit interaction \cite{uma86,kim97,sim01,uma05,mar05,nak05}. In particular, TDHF predictions of barriers and fusion cross sections are in good agreement with experimental data \cite{bon78,sim08,was08,sim13b,sim13c,uma14}. Importantly, it incorporates dynamical effects such as vibration \cite{sim01,obe12} and particle transfer \cite{uma08a,sim10b,sim11} which are crucial at near barrier energies \cite{was08,uma14}.

An important drawback of  mean-field calculations, however, is the impossibility to describe tunneling of the many-body wave function. As a result, sub-barrier fusion cannot be directly investigated in TDHF. Nevertheless, TDHF codes can be used to compute fusion thresholds including dynamical effects \cite{sim04,sim08,was08,guo12,uma14,was15}.
 
To estimate sub-barrier fusion cross sections, one has to determine a nucleus-nucleus potential and to compute the sub-barrier transmission with a barrier penetration model. Microscopic potentials can be calculated directly from TDHF dynamics \cite{uma06b,was08}. Alternatively, one can start with the bare potential computed from a frozen Hartree-Fock approach \cite{sim08,was08} and use a CC code to incorporate the dynamical couplings to collective modes following the approach proposed in Ref.~\cite{sim13c}. The properties of the collective phonons are either known experimentally or computed theoretically, for instance in the linear response theory with a TDHF code \cite{blo79,cho87,sim03,uma05,mar05,nak05,ste07,ave13}. Therefore, the approach of Ref.~\cite{sim13c} allows to use the same Skyrme functional to compute both the bare potential and the deformation parameters of the collective excited states. 

 In the present work, we follow the approach of Ref.~\cite{sim13c} to extract various information on the fusion dynamics:
 \begin{itemize}
 \item The bare potential is computed with the Frozen Hartree-Fock approach.
 \item Fusion cross sections incorporating couplings to low-lying vibrational states are computed with the CC approach using the \textsc{ccfull} code \cite{hag99} with the above potential and experimental properties of low-lying vibrational states (energy and deformation parameters).
 \item The resulting barrier distributions are compared with fusion thresholds computed directly from TDHF, and incorporating dynamical effects at the mean-field level, such as couplings to vibrational states and to transfer channels. 
 \item The importance of transfer channels is estimated with near-barrier TDHF calculations of transfer probabilities using the particle number projection technique developed in Ref.~\cite{sim10b}.
 \end{itemize}
 Combining these information for $^{40}$Ca$+^{58,64}$Ni reactions and comparing with experimental data will then allow to identify the impact of transfer mechanisms on fusion in these reactions. 
 
The HF and TDHF calculations are performed with the \textsc{ev8} \cite{bon05} and \textsc{tdhf3d}  \cite{kim97} codes, respectively. The Sly4$d$ interaction \cite{kim97} is used. Unless specified, pairing is included in static calculations at the BCS level and single-particle occupation numbers are kept constant during the dynamics. The surface pairing interaction is used \cite{ben03}. All calculations are preformed with a mesh space of 0.8~fm and a time step of $1.5\times10^{-24}$~s.

In our calculations, all nuclei are spherical, except for $^{64}$Ni which exhibits a very small triaxial quadrupole deformation with a maximum quadrupole moment of 1.0 fm$^2$. We thus treat it as spherical. Note that other functionals could result in slightly deformed ground-states \cite{dob04,lal99,obe06,sekarXiv}.

\section{Results}
\label{section_Results}

\subsection{Fusion excitation functions}

The frozen Hartree-Fock potential is computed from the HF ground-state one-body density matricies $\rho_{1,2}$ of the two colliding nuclei at a distance $R$ between their centers of mass. The total energy $E[\rho_1+\rho_2](R)$ of the system is given by the energy density functional including Coulomb energy. The nucleus-nucleus potential is then obtained by removing the HF energy of the ground-states $V(R)=E[\rho_1+\rho_2](R)-E_{HF}[\rho_1]-E_{HF}[\rho_2]$.

The resulting bare potential barrier radii $R_b$ and heights $V_b$ are listed in Table~\ref{Frozen_Bass_TDHF_barriers}. As expected, the bare potential barrier radius is larger, and the height smaller, for $^{40}$Ca+$^{64}$Ni than for $^{40}$Ca+$^{58}$Ni due to the larger size of $^{64}$Ni. As a comparison, the bare potential barriers calculated from the frozen HF method and from the Bass model \cite{bas80} are reported in Fig.~\ref{TDHF_barrier_distributions}. The frozen HF method predicts barriers which are noticeably smaller than the more phenomenological parameterisation by Bass. 

\begin{table}[!h]
\centering
\caption{Bare potential barrier characteristics from the frozen HF method and from the Bass model \cite{bas80} as well as TDHF fusion thresholds for $^{40}$Ca+$^{58,64}$Ni.}
\begin{ruledtabular}
	\begin{tabular}{ccc}
		Barrier characteristics & $^{40}$Ca+$^{58}$Ni & $^{40}$Ca+$^{64}$Ni \\
		\hline $V_b^{Frozen}$ [MeV] & 74.15 & 72.26 \\
		\hline $R_b^{Frozen}$ [fm] & 10.11 & 10.40 \\
		\hline $V_b^{Bass}$ [MeV] & 75.73 & 74.66 \\
		\hline $R_b^{Bass}$ [fm] & 10.65 & 10.80 \\
	    \hline $V_b^{TDHF}$ [MeV] & 71.70 & 69.06 \\
	\end{tabular}
\end{ruledtabular}
\label{Frozen_Bass_TDHF_barriers}
\end{table}

\begin{figure}[!ht]
		\centering \includegraphics[width=\columnwidth]{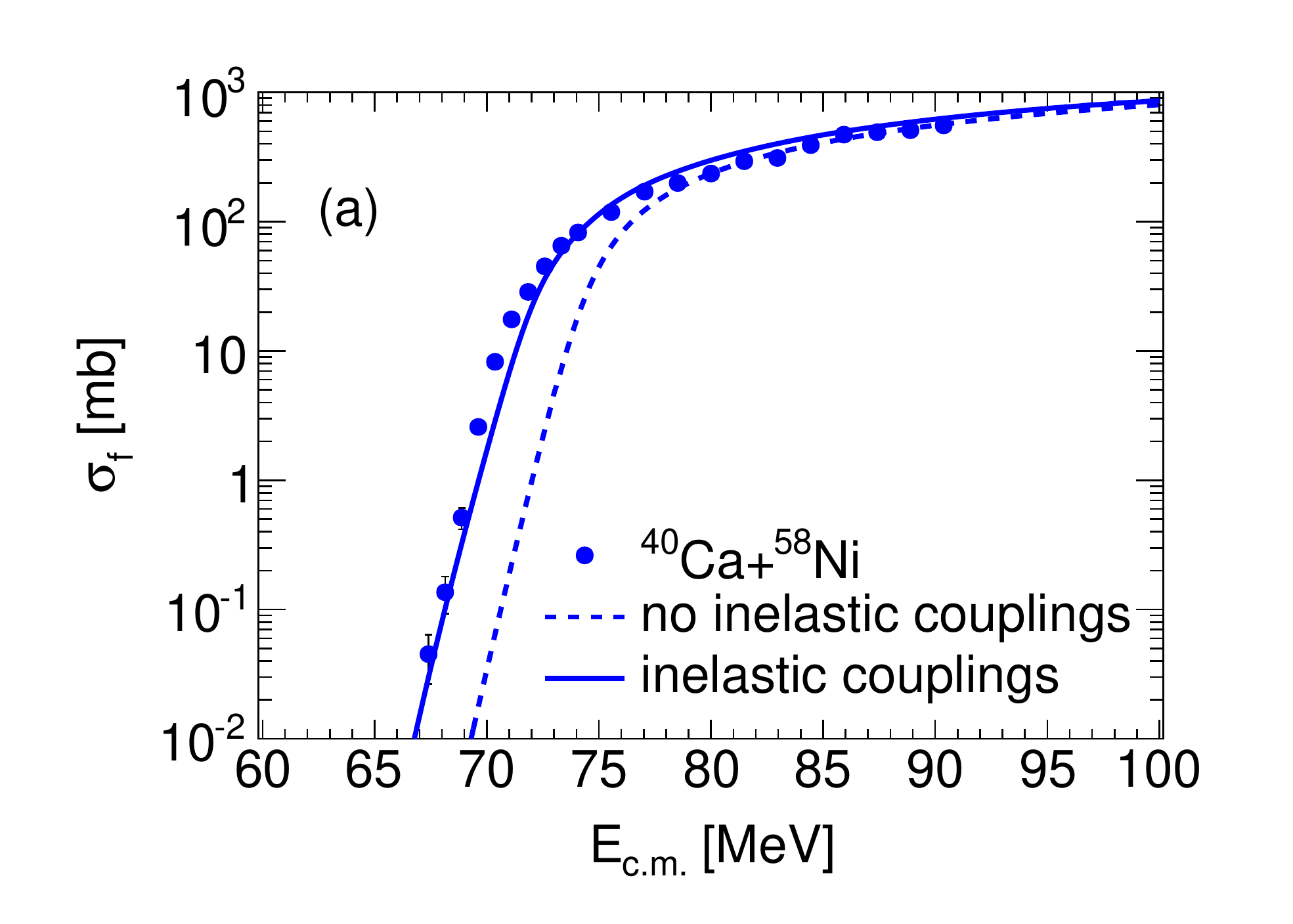}
		\centering \includegraphics[width=\columnwidth]{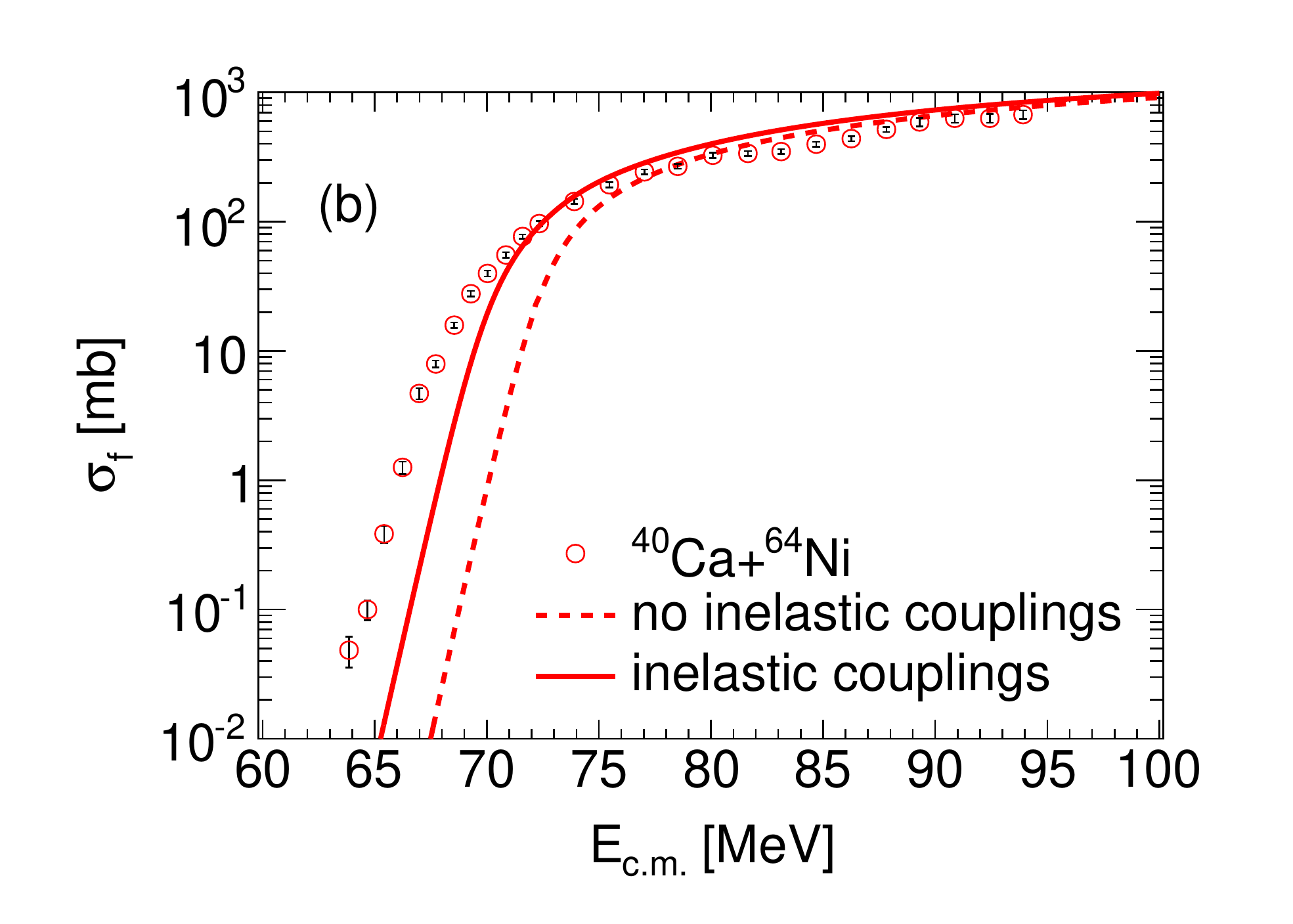}
		\caption{(Color online) Experimental and calculated fusion excitation functions for $^{40}$Ca+$^{58}$Ni (a) and $^{40}$Ca+$^{64}$Ni (b). The coupled-channels calculations were performed with the frozen HF potential and the coupling parameters of the low-lying octupole state for $^{40}$Ca and of the low-lying quadrupole states for $^{58,64}$Ni given in Table~\ref{Deformation_parameters}.}
		\label{FrozenTDHF_fusion_excitation_functions}
\end{figure}

A Saxon-Woods parameterisation of the nuclear potential was extracted from the frozen potential to be used in CC calculations. 
The potential well depths $V_0$, nuclear radii $r_0$ and surface diffuseness $a$ parameters are listed in Table~\ref{SW_parameterisations}. The resulting fusion cross sections $\sigma_f$ were computed with the \textsc{ccfull} code and are plotted with dashed-lines in Fig.~\ref{FrozenTDHF_fusion_excitation_functions}. The experimental data are taken from Ref.~ \cite{bou14}. For the two studied systems, the CC calculations with no couplings to inelastic and transfer channels underestimate the measured fusion cross sections at sub-barrier energies by about two orders of magnitude. 

\begin{table}[!h]
\centering
\caption{Saxon-Woods parameterisation of the nuclear potential extracted from the frozen potential for $^{40}$Ca+$^{58,64}$Ni.}
\begin{ruledtabular}
	\begin{tabular}{cccc}
		System &  $V_0$ [MeV] & $r_0$ [fm] & $a$ [fm] \\
		\hline $^{40}$Ca+$^{58}$Ni &  87.16  &  1.16 & 0.62 \\
		\hline $^{40}$Ca+$^{64}$Ni  & 106.12 & 1.15 & 0.64 \\
	\end{tabular}
\end{ruledtabular}
\label{SW_parameterisations}
\end{table}

The intensity of the couplings to low-lying vibrations is determined by the energy and deformation parameter of the phonons.
The latter can be obtained from nuclear structure experimental data or theoretical calculations. For the present systems, the main effect is expected from the coupling to low-lying octupole modes in $^{40}$Ca and quadrupole modes in $^{58,64}$Ni \cite{hag97,ste95a}. The energy and deformation parameters of the associated first phonons are experimentally well known for these isotopes and are summarised in Table~\ref{Deformation_parameters}.

\begin{table}[!h]
\centering
\caption{Excitation energies ($E$) as well as quadrupole and octupole deformation parameters ($\beta_{2,3}$) of the low-lying vibrational states included in the CC calculations. For references, see the caption of Table I in Ref.~\cite{bou14}.}
\begin{ruledtabular}
	\begin{tabular}{cccc}
		Nucleus & $J^{\pi}$ & $E$ [keV] & $\beta_{2,3}$ \\
		\hline $^{40}$Ca &  $3_1^{-}$  & 3736 & 0.40 \\
		\hline $^{58}$Ni & $2_1^{+}$  & 1454 & 0.18 \\
		\hline $^{64}$Ni & $2_1^{+}$  & 1346 & 0.16 \\
	\end{tabular}
\end{ruledtabular}
\label{Deformation_parameters}
\end{table}

Coupled-channel calculations have been performed including couplings to the low-lying $2^+_1$ and $3^-_1$ states with the \textsc{ccfull} code \cite{hag99}. The resulting fusion cross sections are shown with solid lines in Fig.~ \ref{FrozenTDHF_fusion_excitation_functions}. For $^{40}$Ca+$^{58}$Ni, the measured fusion cross sections are well reproduced over the energy range, especially at sub-barrier energies where the fusion process occurs by quantum tunneling. The couplings to the $3^-_1$ state for $^{40}$Ca and $2^+_1$ state for $^{58}$Ni explain the enhancement of the fusion cross sections in this energy range.

For $^{40}$Ca+$^{64}$Ni, the CC calculations with couplings to inelastic channels still underestimate the measured fusion cross sections at sub-barrier energies. These couplings are the same as those for $^{40}$Ca+$^{58}$Ni since $^{58}$Ni and $^{64}$Ni have similar nuclear properties. Similar results were obtained in Ref.~\cite{bou14}, using the Aky\"uz-Winther nuclear potential and including the same inelastic couplings. However, the diffuseness parameter of the Aky\"uz-Winther nuclear potential was slightly increased for $^{40}$Ca+$^{64}$Ni to fit the fusion cross sections around the Coulomb barrier, whereas the only input parameters in the frozen Hartree-Fock method are those of the Skyrme energy density functional.

Couplings to nucleon transfer channels could also play an important role in the fusion process \cite{jia14}. The $^{40}$Ca+$^{64}$Ni fusion reaction has positive $Q$ values for the transfer of neutrons from $^{64}$Ni to $^{40}$Ca (neutron pick-up) and protons from $^{40}$Ca to $^{64}$Ni (proton stripping), whereas $^{40}$Ca+$^{58}$Ni has only negative nucleon transfer $Q$ values. Corresponding $Q$ values are listed in Table~\ref{Qtransfer}. The importance of neutron transfer in $^{40}$Ca$+^{64}$Ni will be confirmed with TDHF calculations in section~\ref{sec:transfer}.

\begin{table}[!h]
\centering
\caption{Corrected-Q values in MeV of transfer reactions for $^{40}$Ca+$^{58,64}$Ni ($Q_{\rm corr}=Q_{\rm tr}+V^{\rm in}_b-V^{\rm out}_b$) \cite{bro83}. The indicated + sign corresponds to neutron pick-up and the - sign to proton stripping.}
\begin{ruledtabular}
	\begin{tabular}{ccccccc}
		System & +1n & +2n & +3n & -1p & -2p & -3p \\
		\hline $^{40}$Ca+$^{58}$Ni & -3.80 & -2.52 & -11.19 & -3.75 & -3.60 & -11.95  \\
		\hline $^{40}$Ca+$^{64}$Ni & -1.23 & 3.47 & 0.86 & 0.26 & 4.19 & 0.88 \\	
	\end{tabular}
\end{ruledtabular}
	\label{Qtransfer}
\end{table}

\subsection{TDHF fusion thresholds and experimental fusion barrier distributions}
	
Based on a deterministic mean-field approximation, the TDHF theory can only give fusion probabilities which are either 0 or 1 for a given initial energy and impact parameter \cite{bon78}, and for an initial orientation in case of deformed nuclei \cite{sim04}. As a result, it can be used to compute fusion thresholds including dynamical effects due to internal excitations, deformation to all orders, neck formation and nucleon exchange. 

\begin{figure}[!ht]
		\centering \includegraphics[width=\columnwidth]{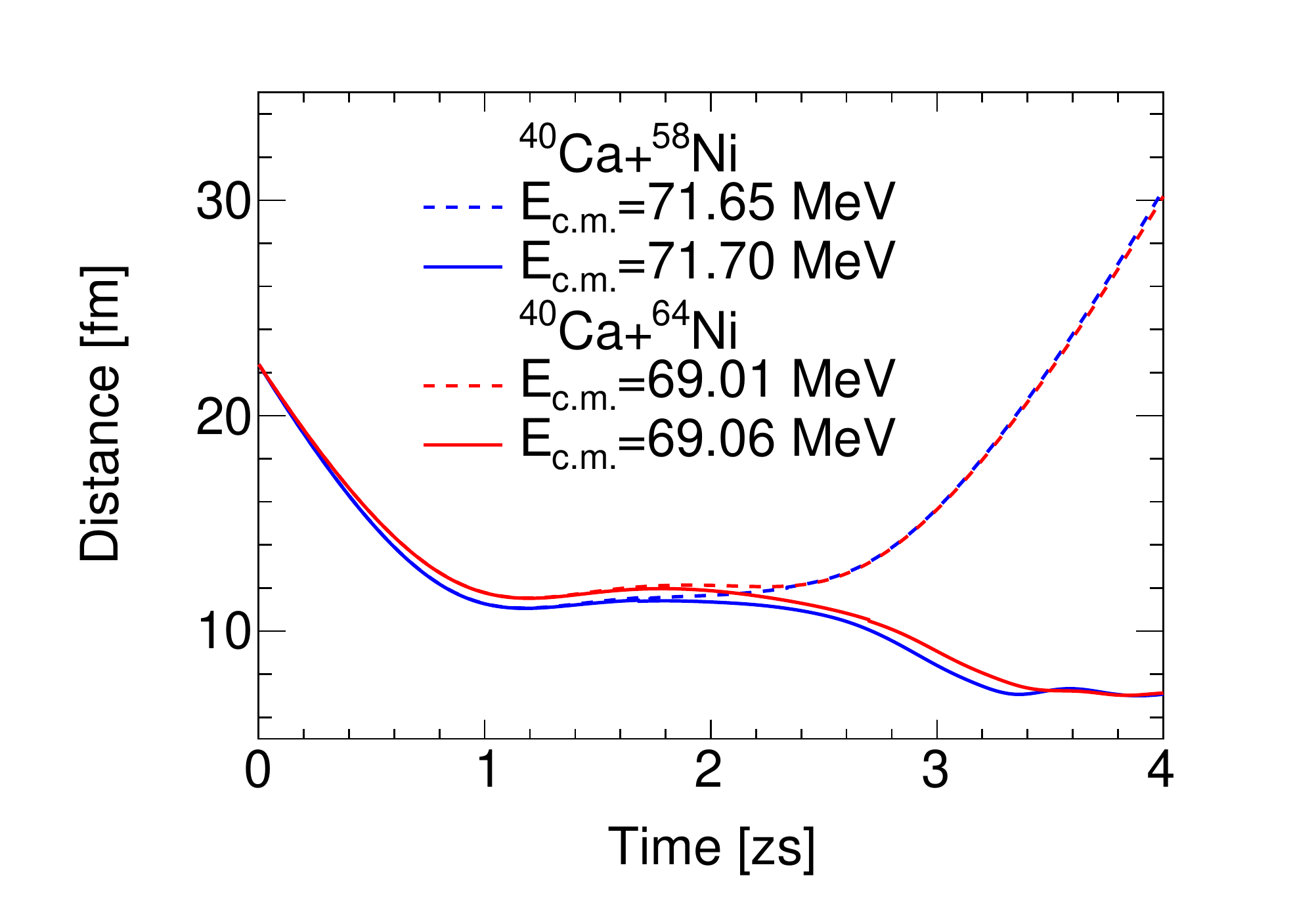}
		\caption{(Color online) Time evolutions of the distance between the colliding nuclei in $^{40}$Ca+$^{58,64}$Ni  just above (solid lines) and below  (dashed lines) the fusion threshold.}
		\label{TDHF_time_distance}
\end{figure}

Examples of  time evolutions of the distance between the colliding nuclei in $^{40}$Ca+$^{58,64}$Ni  are plotted in Fig.~\ref{TDHF_time_distance}. This distance is defined as the distance between the centres of masses computed on each side of the neck. The  energies are chosen to be just above and below the fusion threshold with a difference of only 50~keV.

Interestingly, both systems exhibit similar behaviours. In particular, the fragments are in contact during $\sim1$~zs ($10^{-21}$~s) before fusion is decided. Similar times were obtained for  near barrier collisions in lighter \cite{kim97,leb12} and  similar \cite{sim11,sim13c} mass regions. These times are relatively short in comparison with heavier systems computed with TDHF where typical contact times of few zeptoseconds \cite{sim12a} or more in case of quasi-fission reactions \cite{guo12,sim12b,wak14,obe14,uma15a} are often obtained, leading to possible large mass transfer \cite{tok85,rie11}. Nevertheless, contact times of $\sim1$~zs are long enough to allow transfer of one or more nucleons \cite{koo77,sim10b,yil11,yil14}, in particular in the case of positive $Q$ value reactions as will be seen in section~\ref{sec:transfer}.

The resulting fusion thresholds are listed in Table~\ref{Frozen_Bass_TDHF_barriers}. These thresholds are lower than the bare potential barrier heights computed with the frozen HF method (2.45 MeV for $^{40}$Ca+$^{58}$Ni and 3.20 MeV for $^{40}$Ca+$^{64}$Ni). This lowering of the fusion barrier height is induced by dynamical effects such as couplings to inelastic and nucleon transfer channels. These couplings are also responsible for the enhancement of sub-barrier fusion cross sections with respect to the uncoupled case (see Fig.~\ref{FrozenTDHF_fusion_excitation_functions}). 

\begin{figure}[!ht]
		\centering \includegraphics[width=\columnwidth]{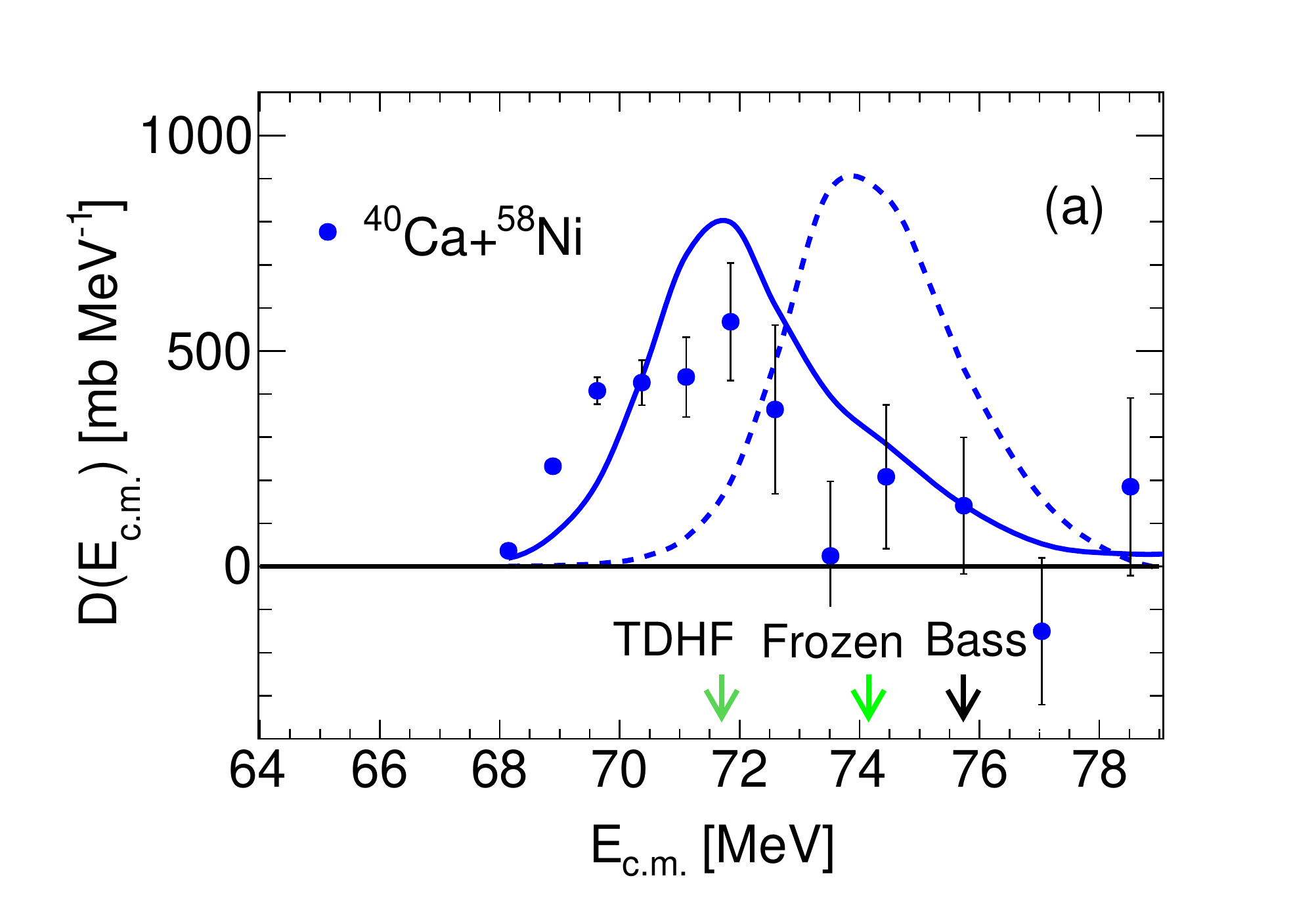}
		\centering \includegraphics[width=\columnwidth]{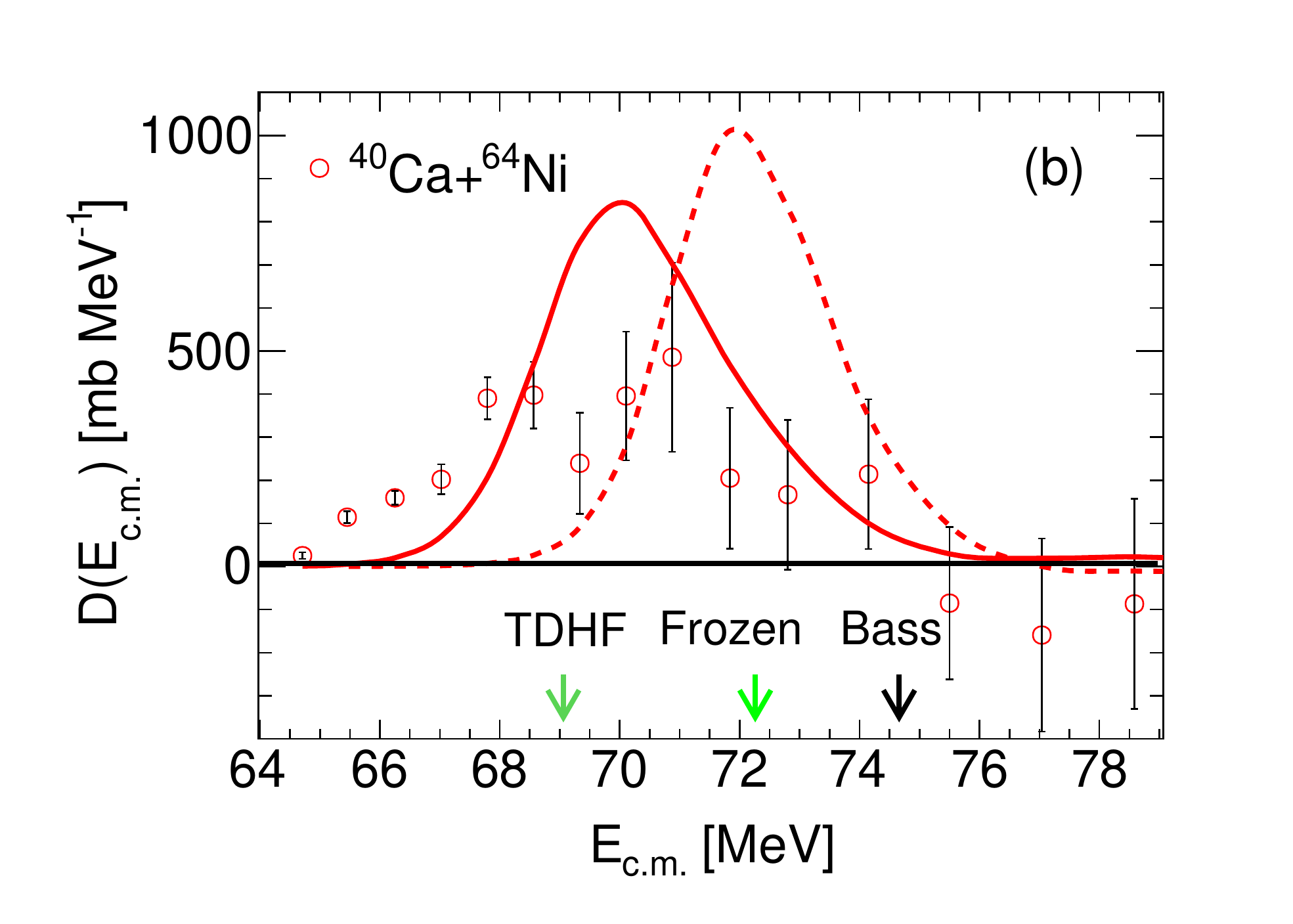}
		\caption{(Color online) Experimental and calculated barrier distributions for $^{40}$Ca+$^{58}$Ni (a) and $^{40}$Ca+$^{64}$Ni (b). The coupled-channels calculations were performed with the \textsc{ccfull} code, using the frozen potential and the coupling parameters of the low-lying octupole state for $^{40}$Ca and of the low-lying quadrupole states for $^{58,64}$Ni given in Table~\ref{Deformation_parameters}. The Bass, frozen HF and TDHF barriers are shown with arrows on the $E_{c.m.}$-axis.}
		\label{TDHF_barrier_distributions}
\end{figure}

The fusion and bare potential barrier heights can be compared to the centroids of the measured and calculated barrier distributions which are plotted in Fig.~\ref{TDHF_barrier_distributions}. These barrier distributions \cite{row91} were derived from the second derivative of $(E_{c.m.} \times \sigma_f)$ with respect to $E_{c.m.}$, using the three-point difference formula \cite{das98}, with an energy step of $\Delta E_{c.m.} \simeq 1.5$ MeV below the Coulomb barrier and 3 MeV above the Coulomb barrier. 

We see in Fig.~\ref{TDHF_barrier_distributions} that the TDHF fusion threshold for $^{40}$Ca+$^{58}$Ni is in good agreement with the large structure of the measured barrier distribution as well as the maximum of the CC barrier distribution (solid line). 
As the TDHF result includes all type of couplings, this indicates that most of the lowering of the fusion threshold is accounted for by the coupling to the $3^-_1$ state in $^{40}$Ca and to the  $2^+_1$ state in $^{58}$Ni.

For $^{40}$Ca+$^{64}$Ni, the TDHF barrier is also in agreement with the large structure of the measured barrier distribution but it is lower than the maximum of the CC barrier distribution (solid line). This indicates that the couplings to transfer channels, which are included in TDHF dynamics but not in the CC calculations, could play a role in the lowering of the fusion threshold. 

\subsection{Sub-barrier transfer probabilities}
\label{sec:transfer}

Neutron transfer probabilities were computed using the particle number projection technique developed in Ref.~\cite{sim10b}.
For simplicity, these calculations were performed without pairing correlations. This is sufficient to provide a reasonable estimate of the overall importance of transfer. More predictive transfer probabilities for individual transfer channels would require the inclusion of dynamical pairing correlations \cite{ave08,sca13}.

Neutron number distributions of the Ca projectile-like fragment in the exit channel were calculated at $E_{c.m.}=71.65$ MeV for $^{40}$Ca+$^{58}$Ni and at $E_{c.m.}=69.01$ MeV for $^{40}$Ca+$^{64}$Ni. These energies are located just below the TDHF fusion thresholds.

\begin{figure}[!ht]
		\centering \includegraphics[width=\columnwidth]{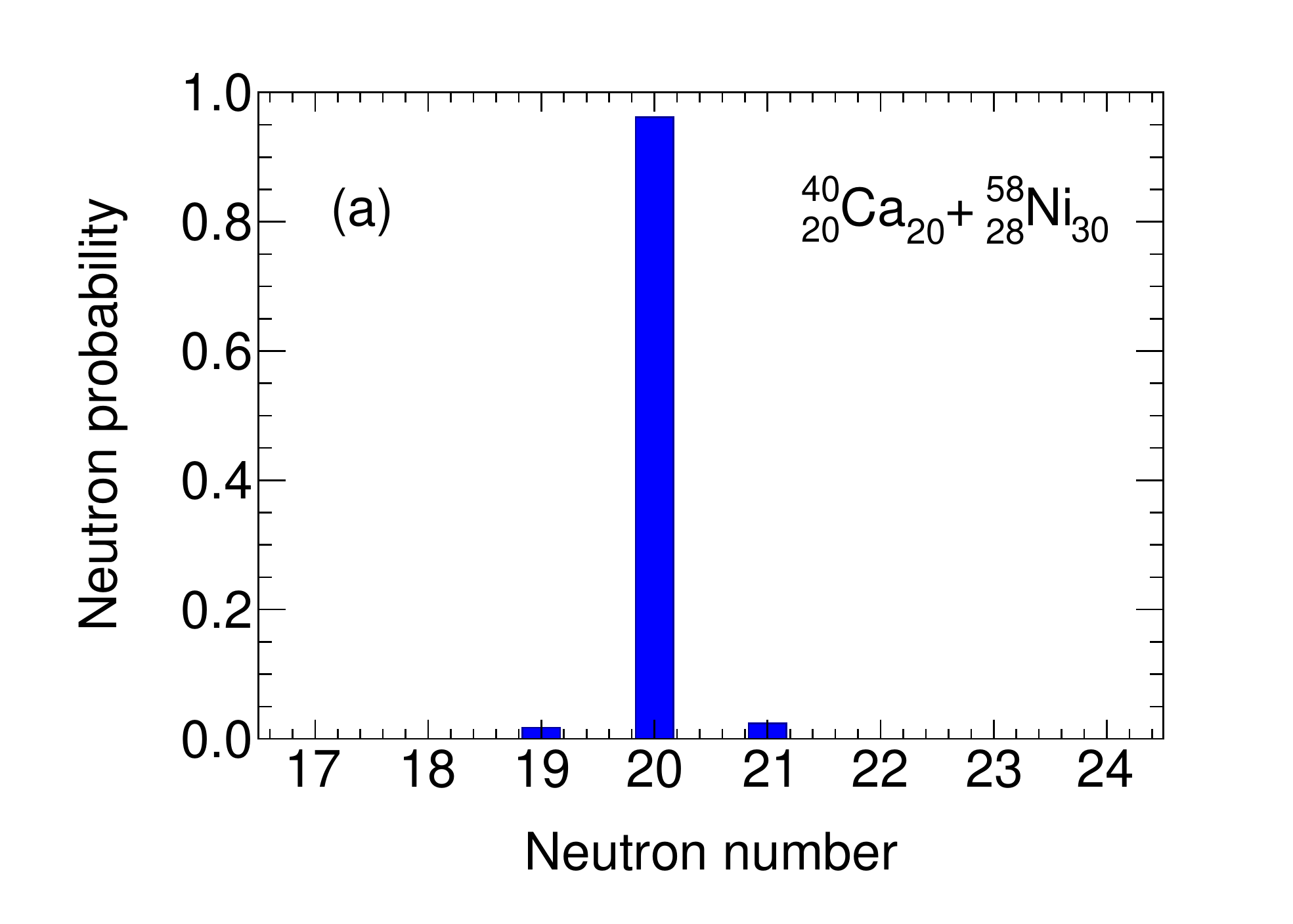}
		\centering \includegraphics[width=\columnwidth]{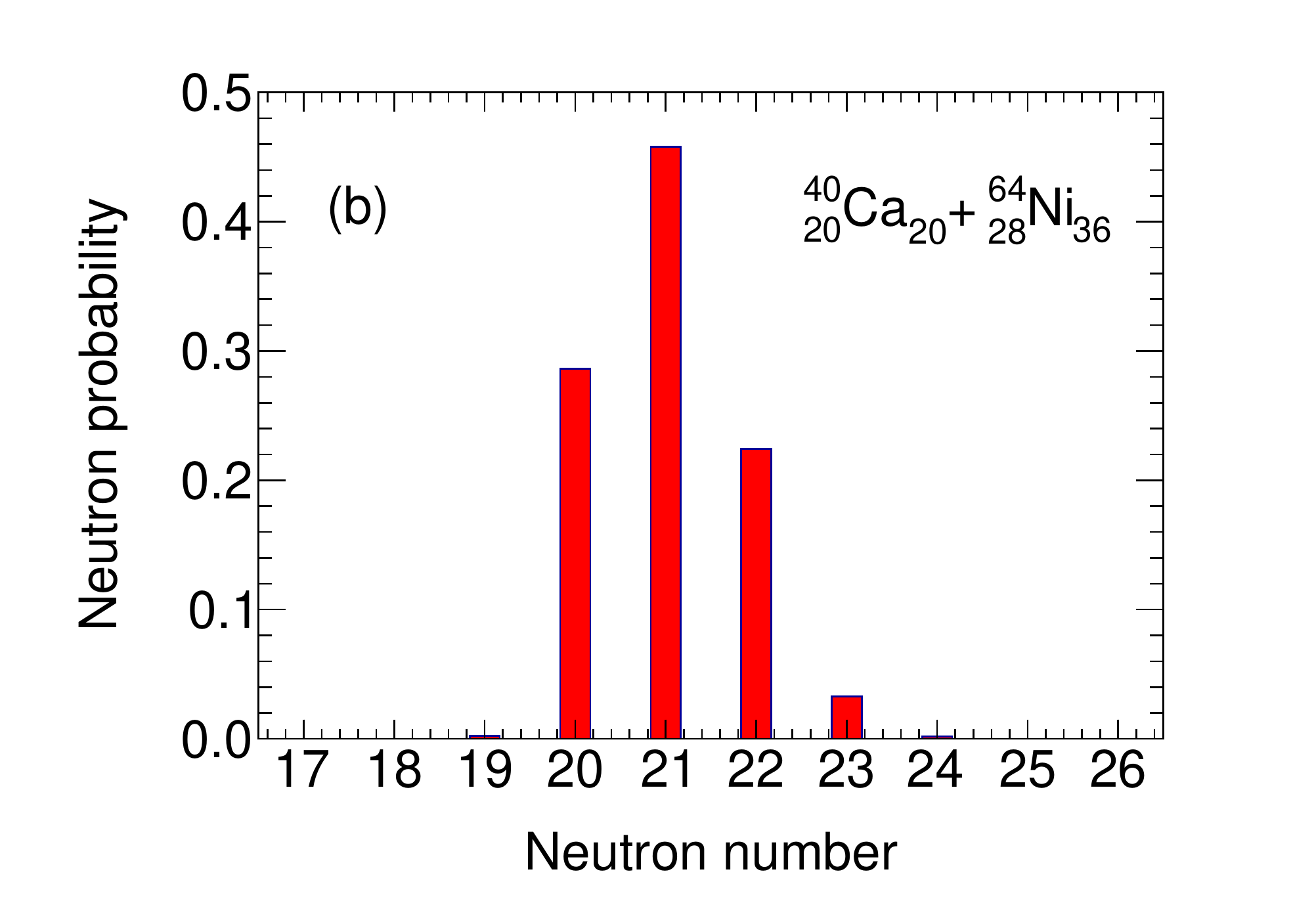}
		\caption{(Color online) TDHF calculations of the Ca quasi-projectile neutron number distributions in the exit channel of (a) $^{40}$Ca+$^{58}$Ni at $E_{c.m.}=71.65$ MeV and (b) $^{40}$Ca+$^{64}$Ni at $E_{c.m.}=69.01$ MeV.}
		\label{TDHF_transfer}
\end{figure}

As can be seen in Fig.~\ref{TDHF_transfer}, the sum of the transfer probabilities does not exceed $0.04$ in $^{40}$Ca+$^{58}$Ni near barrier collisions. This confirms our previous conclusion that the impact of transfer channels on fusion is small in this reaction. This was expected due to negative $Q$ values for transfer reactions in this system.

Transfer probabilities are much larger in the $^{40}$Ca+$^{64}$Ni system, with a total transfer probability of $\sim0.7$. 
In this case, the dominant channel is the transfer of one neutron  from $^{64}$Ni to $^{40}$Ca, followed by the two-neutron transfer channel. However, the relative weighting between individual channels could be strongly modified with the inclusion of dynamical pairing \cite{sca13} favouring pair transfer channels \cite{oer01,mon14}.

Quantitatively, the couplings to the vibrational states accounts for a lowering of the barrier by $\sim2$~MeV in  $^{40}$Ca+$^{64}$Ni, while TDHF calculations predicts that the fusion thresholds should be lowered by more than 3~MeV. 
It is likely that the transfer channels are responsible for this difference. Indeed, they could easily affect the fusion process by, e.g.,  facilitating  the formation of a neck between the fragments. 

\section{Summary} 
\label{section_Summary}

Microscopic and coupled-channel calculations have been performed for the  $^{40}$Ca+$^{58,64}$Ni reactions near the fusion barrier. The only input parameters are those of the Skyrme energy density functional. Fusion cross sections around and below the Coulomb barrier were obtained from CC calculations, using the bare nucleus-nucleus potential computed with the frozen Hartree-Fock method and coupling parameters taken from known nuclear structure data. Couplings to the $3^-_1$ state in $^{40}$Ca and to the $2^+_1$ states in $^{58,64}$Ni were included in the CC calculations. The resulting fusion barrier distributions were compared with experimental data as well as TDHF fusion thresholds which include automatically all types of couplings (inelastic and nucleon transfer channels) at the mean field level. 

The lowering of the fusion threshold due to dynamical effects is explained in $^{40}$Ca+$^{58}$Ni without invoking transfer channels which are shown to be weak. The case of $^{40}$Ca+$^{64}$Ni is different as inelastic couplings to low-lying phonons only account for about two third of the lowering of the fusion thresholds. Transfer channels which are shown to be important near barrier, could result in a further lowering of the fusion thresholds, possibly via neck formation. 

Experimental measurement of nucleon transfer cross section around the fusion barrier for these two systems were recently performed to investigate directly this effect. It would also be interesting to extend the measurement of fusion cross sections at lower energies for these systems in order to better understand the role of transfer reactions on the deep sub-barrier fusion hindrance recently observed \cite{jia04,das07}.

Further theoretical investigations are also required to better understand the mechanisms of the couplings between transfer channels and fusion. In particular, transfer could favour neck formation and, thus, diffusion towards a compound nucleus. It has also been argued that transfer could be a doorway to dissipation hindering fusion \cite{eve11}.

\section{Acknowledgments}

The research leading to the experimental results mentioned in this paper has received funding from the European Commission within its Seventh Framework Programme under Grant Agreement No. 262010 ENSAR. This work has been supported by the Australian Research Council Grant No. FT120100760.

\bibliographystyle{unsrt}

\end{document}